\documentstyle[12pt]{article}

\begin{document}

\title{{\small UPR-0126MT\hfill Dec. 1992}\\
Dynamics of antibaryon-baryon annihilation in the
Skyrme model}

\author{
Bin Shao, Niels R. Walet, and R. D. Amado\\
Department of Physics, University of Pennsylvania,\\
Philadelphia, PA 19104}

\date{December 18, 1992}
\maketitle

\begin{abstract}
We examine the dynamics of a baryon number zero lump in
the Skyrme picture, as a model for annihilation in the $\bar{N}N$ system.
We find that radiation propagates at the causal limit
as a localized pulse and that
the linearized theory gives a good approximation. Both of these
results are contrary to
findings in the Sine-Gordon model.
\end{abstract}

\newpage

Well before the advent of QCD, Skyrme \cite{Skyrme}
proposed  a non-linear
meson field theory to describe baryons in terms of  soliton
solutions with nontrivial topologies. He identified the conserved
topological charge with the baryon number $B$ and thus built in
the conservation law for the baryon number in an elegant way.
This model is now called the Skyrme model. In the Skyrme model,
a classical solution of the field equation with $B=1$ is used to
describe a nucleon and a solution with $B=2$ is used for a system of
two nucleons. Similarly a solution with $B=0$ can be used to study
a antinucleon-nucleon system. There has been a vast amount of work
on $B=1$ solutions to study static properties of nucleons \cite{Reviewpapers}
 and on
$B=2$ solutions to study nucleon-nucleon interactions \cite{NN}. The
Skyrme model has been shown to provide a viable phenomenological
description of nucleons and their interactions. In contrast, much less
attention has been paid to  $B=0$ solutions. Recently Sommermann
{\em et al} \cite{Seki} performed a
comprehensive calculation for a $B=0$ solution
by integrating the classical equations of motion on a
3D grid numerically. Many interesting features in the time evolution of the
antiskyrmion-skyrmion system were observed. For example, the
baryon density is found to decay extremely rapidly upon collision
of the skyrmion with the antiskyrmion, close to the causal limit,
while the energy distribution remains concentrated in the annihilation region.
No satisfactory explanation for the remarkably rapid disappearance
of baryon number is offered by Sommermann {\em et al}.
The full calculation of an antiskyrmion-skyrmion collision is
complicated and it is difficult to extract simple dynamical insight
from it. In this paper we construct a greatly simplified system to explore
the dynamics. We find that a spherically
symmetric ``lump''  with $B=0$
decays quickly, nearly at the causal limit. We also find,
surprisingly, that this decay is qualitatively  described by
the linearized Skyrme equation. By contrast, we find that for the 1+1
Sine-Gordon (SG)
equation there is very slow radiative  decay of a localized $B=0$
``lump''. The lump sits ``ringing'' for a long time, slowly radiating its
energy away. The linearized SG case radiates its energy more
rapidly, but not nearly as quickly as the skyrmion case.

\begin{figure}
\epsfysize=6cm
\centerline{\epsffile{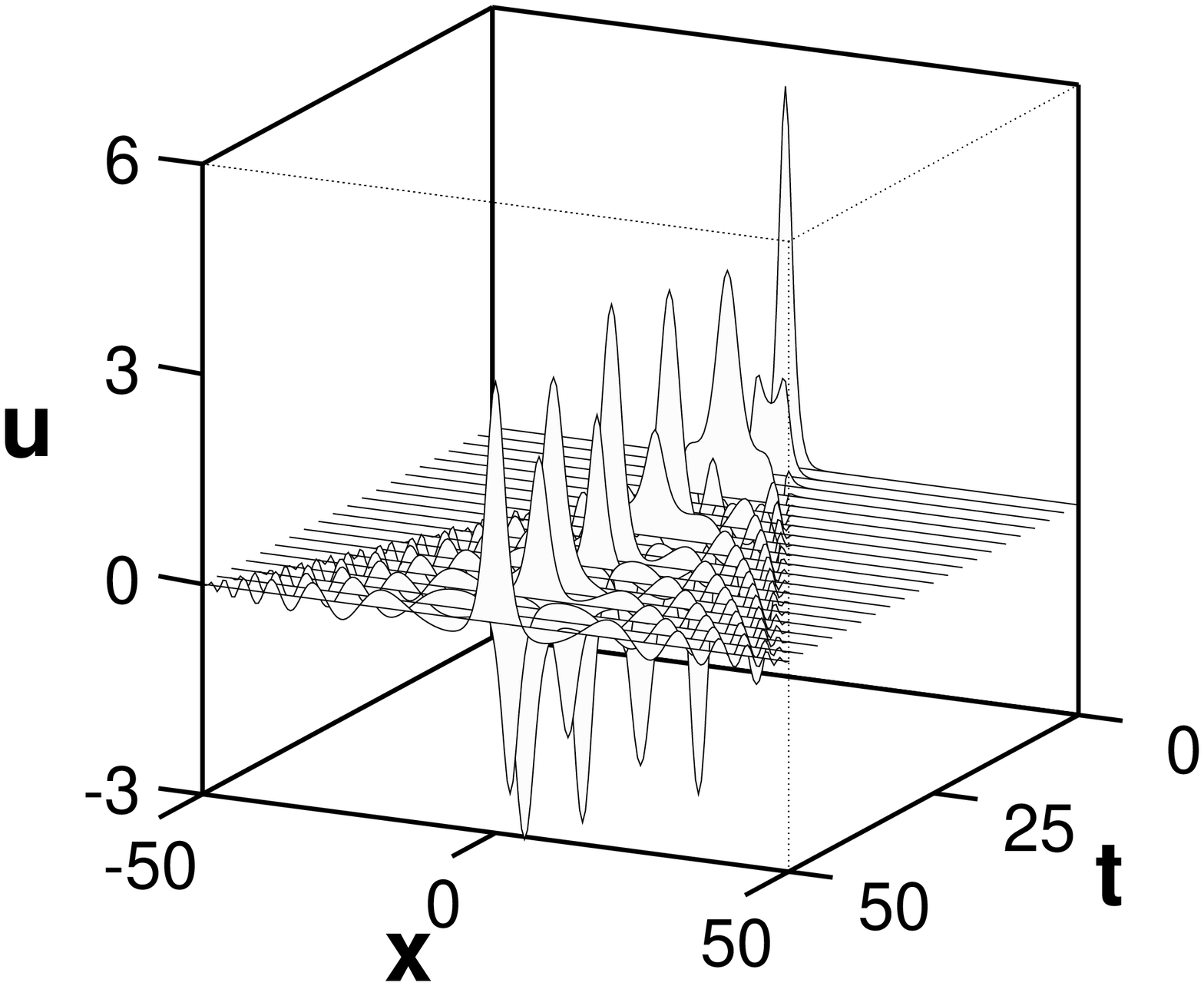}}
\caption{Time evolution of the amplitudes of the solution of the SG
equation (\protect{\ref{eq:SG}})
with initial values (\protect{\ref{eq:SGinit1}}).}
\end{figure}
\begin{figure}
\epsfysize=6cm
\centerline{\epsffile{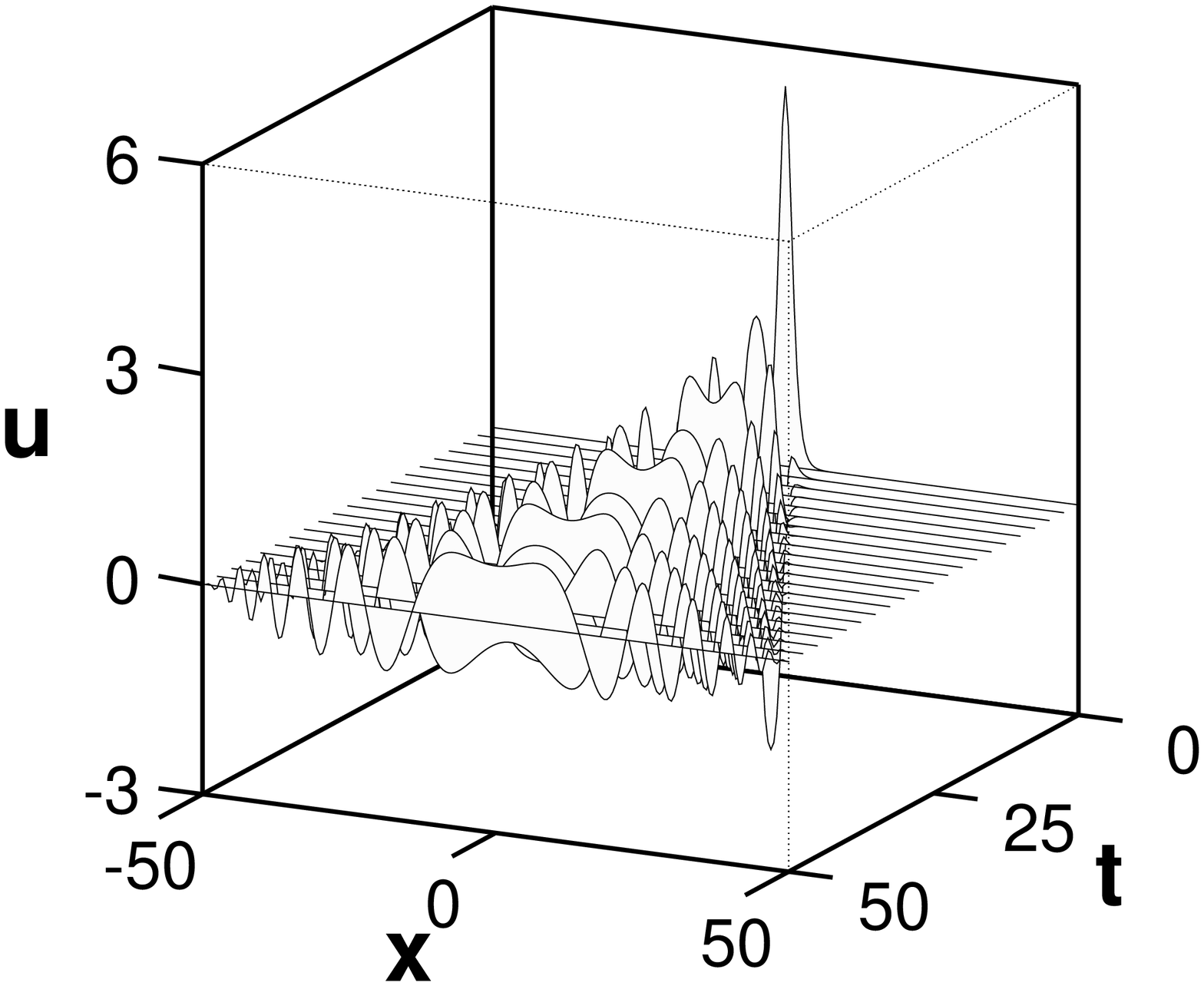}}
\caption{Time evolution of the amplitudes of the solution of the linear wave
equation (\protect{\ref{eq:SGlinear}})
with initial values (\protect{\ref{eq:SGinit1}}).}
\end{figure}

We start with the 1+1 SG model \cite{SineGordon}. As a nonlinear
field theory itself, the SG model has often been used
as a toy model for skyrmions \cite{Toy}. In fact, soon
after the Skyrme model was proposed, Skyrme together with Perring \cite{PS}
studied time evolution of a two soliton system numerically using
the SG model.
The equation of motion for the SG model is
\begin{equation}
\frac{\partial^{2}u}{\partial t^{2}}-
\frac{\partial^{2}u}{\partial x^{2}}+\sin u=0.  \label{eq:SG}
\end{equation}
In linear approximation, this equation becomes
\begin{equation}
\frac{\partial^{2}u}{\partial t^{2}}-
\frac{\partial^{2}u}{\partial x^{2}}+u=0. \label{eq:SGlinear}
\end{equation}
The topological conserved ``baryon number'' for the SG model
is $B=u(x=\infty)-u(x=-\infty)$. To study the
 decay of a $B=0$ system we take
the initial configuration
\begin{equation}
u(0,x)=h\frac{aK_{1}(\sqrt{x^{2}+a^{2}})}{\sqrt{x^{2}+a^{2}}},
\label{eq:SGinit1}
\end{equation}
with $\frac{\partial u}{\partial t}(0,x)=0$.
This choice allows us to solve the linearized equation (\ref{eq:SGlinear})
 analytically.  It represents
 a $B=0$ lump concentrated around the origin with zero
initial velocity.
The parameters are chosen such that the lump has
twice the energy  of a $B=1$ soliton and thus it represents
a $\bar{N}N$ system at threshold. We take h=39 and a=2.
In Figures 1 and 2
we show the time evolution of the $B=0$ solution $u(x)$ for the exact
and linear SG equations, respectively.
As we can see, the solution to the exact SG equation
oscillates very strongly but remains concentrated around the origin
and there is very little dissipation.
This may be related to the infinite number of conservation laws
in the SG model.
For the linear solution, there is more radiation. However,
some portion of the energy also remains around the origin. Apparently,
the linear solution is  different from the exact solution. This is
not surprising since $u$ is rather large initially and the approximation
$\sin u\approx u$ does not apply. Nevertheless both the linear and
full equations share the feature that the distribution sits ``ringing''
near the origin for a long time slowly radiating away its energy.

\begin{figure}
\epsfysize=6cm
\centerline{\epsffile{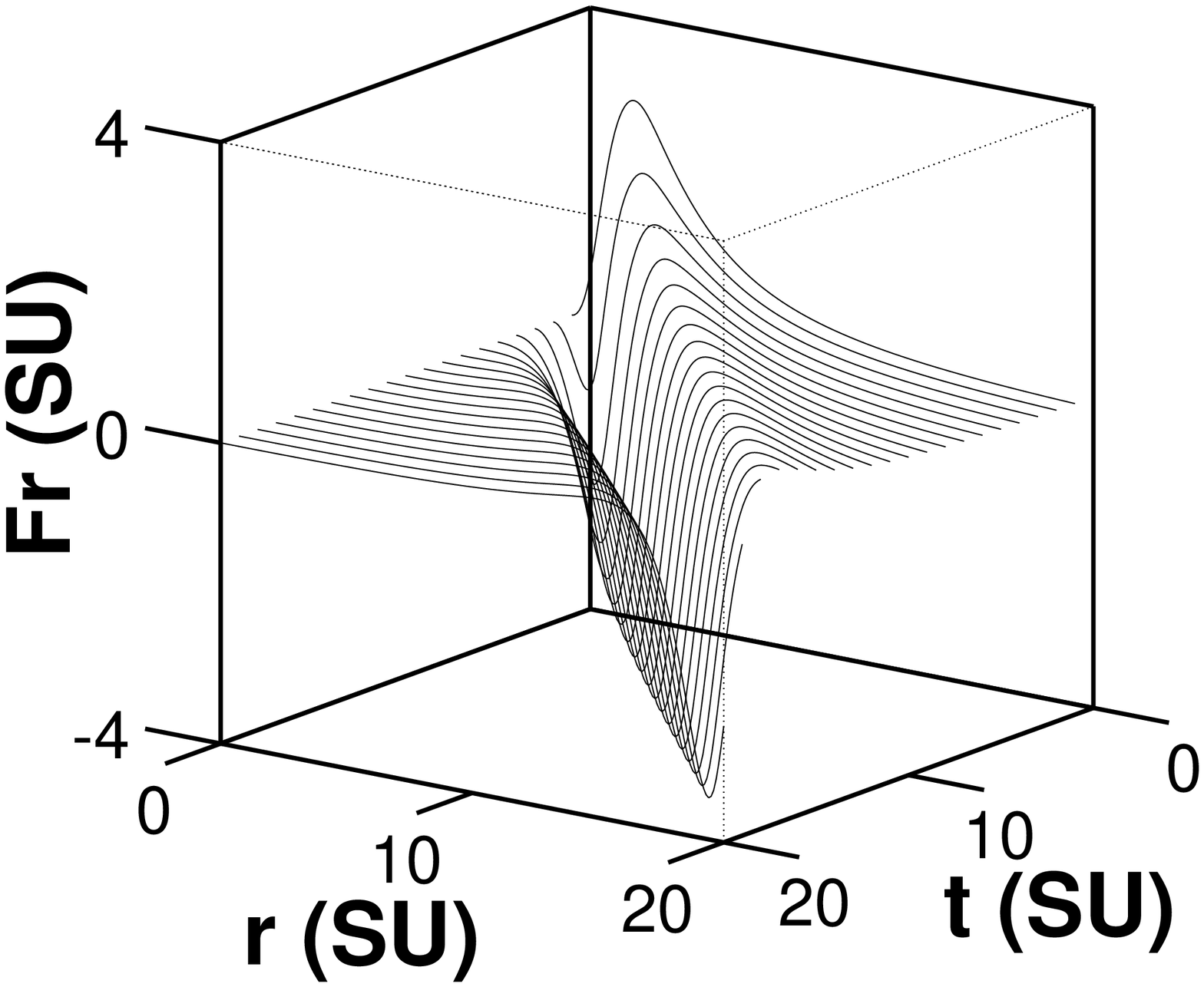}}
\caption{Time evolution of the amplitude (multiplied by $r$) of the solution
of the Skyrme equation (\protect{\ref{eq:SK}})
with initial values (\protect{\ref{eq:SKinit1}}).}
\end{figure}
\begin{figure}
\epsfysize=6cm
\centerline{\epsffile{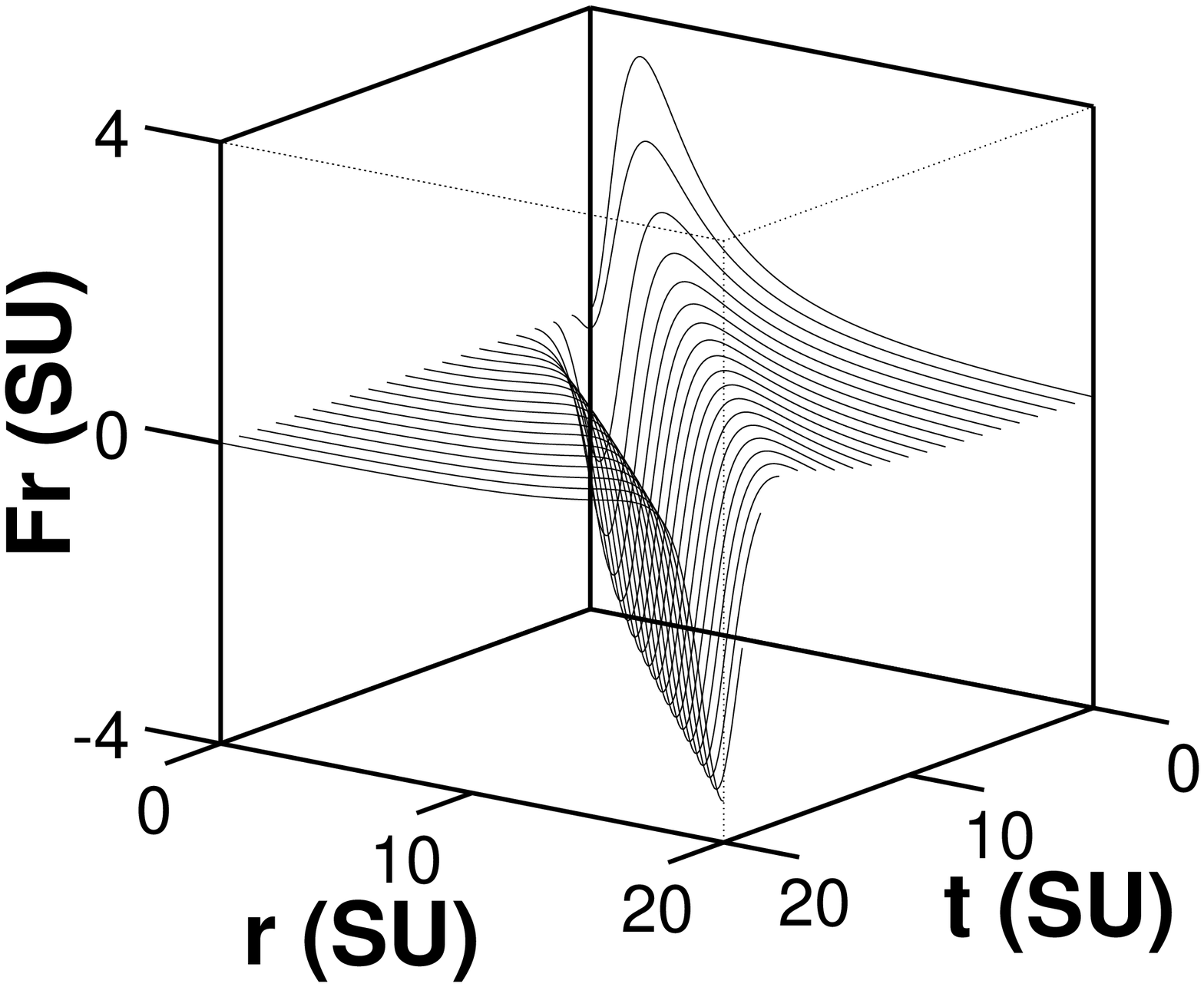}}
\caption{Time evolution of the amplitude (multiplied by $r$) of the solution
of the linear $P$-wave equation (\protect{\ref{eq:SKlinear}})
with initial values (\protect{\ref{eq:SKinit1}})}
\end{figure}

Naively, we would expect to see similar features in the Skyrme
model. This turns out not to be the case.  The Lagrangian density of the
Skyrme model with zero pion mass is
\begin{equation}
{\cal L}=\frac{f^{2}_{\pi}}{4}Tr(\partial_{\mu}U\partial^{\mu}U^{\dagger})
+\frac{1}{32g^{2}_{\rho}}
Tr[U^{\dagger}\partial_{\mu}U,U^{\dagger}\partial_{\nu}U]
[U^{\dagger}\partial^{\mu}U,U^{\dagger}\partial^{\nu}U],
\end{equation}
where $U$ is a unitary $SU(2)$ matrix normalized to $U=1$ at infinity.
If we use the hedgehog ansatz for $U$
\begin{equation}
U=\exp(i\hat{r}\cdot{\tau}F(r,t)),
\end{equation}
we obtain the equation of motion
\begin{equation}
(r^{2}+2\sin^{2}F)(\frac{\partial^{2}F}{\partial t^{2}}
-\frac{\partial^{2}F}{\partial r^{2}})=
2r\frac{\partial F}{\partial r}-\sin 2F[1+\left(\frac{\partial F}
{\partial t}\right)^{2}
-\left(\frac{\partial F}
{\partial r}\right)^{2}
+\frac{\sin^{2}F}{r^{2}}]. \label{eq:SK}
\end{equation}
In arriving at the  above equation, we
used scaled Skyrme units for length and time:
$\tilde{r}=f_{\pi}g_{\rho}r,\tilde{t}=f_{\pi}g_{\rho}t$.
For convenience, we still use $r$ and $t$ to denote these scaled quantities.
In the linear approximation, this equation becomes
\begin{equation}
\frac{\partial^{2}F}{\partial t^{2}}
-\frac{\partial^{2}F}{\partial r^{2}}-
\frac{2}{r}\frac{\partial F}{\partial r}
+\frac{2}{r^{2}}F=0. \label{eq:SKlinear}
\end{equation}
The boundary values  of $F$ are connected  to the baryon number $B$ by
\begin{equation}
B=\frac{1}{\pi}(F(0,t)-F(\infty,t)).
\label{eq:Bsk}
\end{equation}
To model the $\bar{N}N$ system with $B=0$, we take
as an  initial configuration
\begin{equation}
F(r,0)=h\frac{r}{(r^{2}+a^{2})^{2}},
\label{eq:SKinit1}
\end{equation}
with $\frac{\partial F}{\partial t}(r,0)=0$.
Again this particular initial condition makes it possible to solve
the equation of motion in the linearized model analytically.
We choose  parameters such that the energy of the lump
is twice the energy of a single skyrmion. We take $h=49.6$ and $a=2$.
Eq. (\ref{eq:SK}) is solved by an implicit
difference scheme. The amplitudes $F(r,t)$ (multiplied
by $r$) are shown in Figures 3-4.
The most
striking feature we see is that there is no radiation in either case.
Both solutions propagate as a localized pulse, essentially at the
speed of light. Moreover, despite the fact that
the initial lump has rather large $F$ and thus invalidates the
linear approximation, the linear solution is very close to the
exact solution. This implies that in
the Skyrme model, decay processes of an $\bar{N}N$ system can be  described
by a linear meson theory, at least classically. We have  performed
similar calculations with nonzero pion mass and different initial
configurations. We always find very similar solutions for the linear
and non-linear equations and find very little radiation. Rather the
disturbance propagates away from the origin nearly at the causal limit
as seen by Sommermann {\em et al}. They also find little sensitivity to
the pion mass.

That the SG solutions and the Skyrme
solutions exhibit very different behaviors should serve as a
strong warning that qualitative conclusions one draws from studying
the 1+1 SG  model do not necessarily apply to the Skyrme model.
There are at least two reasons one can
give for this difference. First at the level of the linear equation,
the difference seen between the skyrmion and SG cases is largely
due to the meson mass.
In the SG
case the mass in dimensionless unit is 1 since it is the only scale in
the problem. One can define a massless equation of the linear form, but
a massless SG model makes no sense. By contrast in the Skyrme model
the mass scale is set by $f_{\pi}$
and $g_{\rho}$. In those units the pion mass squared is only 0.27.
That is why one sees coherent  propagation in the linearized Skyrme
case and dispersive breakup of the lump in the linearized SG
case. The  smallness of the pion mass in skyrmion units also
accounts for
the lack of sensitivity to the addition of this mass term to the Skyrme
model that we  and  Sommermann {\em et al}
 find. The  dispersive breakup of the SG lump in the
full equation is partly due to this mass effect and partly due to the
intrinsic nonlinearities of the theory.  These nonlinearities
are  much less important in the Skyrme case than in the SG case
due to
the different geometry.
It is easy to show that the nonlinear term in the Skyrme
case (the difference between (\ref{eq:SK}) and (\ref{eq:SKlinear}) )
falls off like $1/r^{2}$ due to the  three dimensional
nature of the equation. Hence, no matter how big the nonlinear terms are
near the origin, they fall quickly and the linear approximation becomes
good.  The boundary on the Skyrme equation, Eq. (\ref{eq:Bsk}), requires
that $F$ be zero at the
origin, further favoring large $r$ and therefore the domain of the linear
equation.
It is clear therefore that the rapid evaporation
of the energy in a $B=0$ Skyrme system is a rather general feature of these
classical models. The robust nature of the rapid annihilation and its
insensitivity to input parameters (recall we are using scaled variables)
has also been observed by Sommermann {\em et al}.

Our results lead to the conclusion that annihilation in the Skyrme model is
described by a rapid decay and that the linearized equation describes the
evolution of this decay very well. These results may have  bearing on
the  quantum-mechanical annihilation in the antinucleon-nucleon system.
First of all it is found experimentally \cite{PDT} that near threshold
the $\bar{N}N$  system couples strongly to purely pionic states. Thus
the Skyrme model, which is formulated in terms of pions only,
 may be a good paradigm for  annihilation in this system.
The Skyrme picture models QCD in
the large number-of-colors ($N_C$) or mean-field limit. From our
results we cannot yet conclude how to describe the formation of pions from
the initial $\bar{N}N$ state.
However formed, we can conclude that the state should disappear rapidly,
nearly at the causal limit.
 The validity of the linear approximation
suggests that after formation  we can use a
model of non-interacting pions (the quantized from of the linear equation)
for the time evolution of the system.
The fast decay implies high kinetic energy of the pions, and we thus
expect that the total number of pions produced is small compared to
$2~{\rm GeV}/m_\pi \approx 14$, as born out by experiment \cite{PDT}.
Such a picture in terms of
rapid decay into a small number of relativistic, non-interacting
pions is very far from the thermodynamic limit.

We thank Prof. R. Seki for interesting discussions.
This work is supported in part by grants from the National Science
Foundation.


\begin{thebibliography}{99}
\bibitem{Skyrme}
T. H. R. Skyrme, Proc. Roy. Soc. London {\bf 260} (1961) 127;
{\bf 262} (1961) 237;
Nucl. Phys. {\bf 31} (1962) 556.
\bibitem{Reviewpapers}
I. Zahed and G. E. Brown, Phys. Rep {\bf 142} (186) 1;
K. F. Liu (ed.), {\em Chiral Solitons}, (World Scientific, Singapore, 1987),
and references therein.
\bibitem{NN}
N. R. Walet {\em et al}, 
Phys. Rev. Lett. {\bf 68} (1992) 3849;
N. R. Walet and R. D. Amado, Phys. Rev. C in press;
A. Jackson {\em et al}, 
Nucl. Phys. {\bf A432} (1985)
567; R. Vinh Mau {\em et al},
Phys. Lett. {\bf B150} (1985) 259; T. S. Walhout
and J. Wambach, Phys. Rev. Lett. {\bf 67} (1991) 314.
\bibitem{Seki}
H. M. Sommermann {\em et al},
Phys. Rev. {\bf D45} (1992) 4303.
\bibitem{SineGordon}
R. Rajaraman, {\em Solitons and Instantons}, (North-Holland,
Amsterdam, 1989);
M. J. Ablowitz {\em et al},
Siam J. Appl. Math., {\bf 36} (1979) 428-437.
\bibitem{Toy}
R. D. Amado {\em et al},
Phys. Rev. Lett. {\bf 63} (1989) 852;
M. Oka {\em et al},
Phys. Rev. {\bf C39} (1989) 2317.
\bibitem{PS}
J. K. Perring and T. H. R. Skyrme,
Nucl. Phys., {\bf 31} (1962) 550-555.
\bibitem{PDT} J. Sedi\'ak and V. \^Simak, Sov. J. Part. Nucl. {\bf 19}
(1988) 191.
\end{thebibliography}
\end{document}